\documentclass[creativecommons]{eptcs}

\usepackage{graphicx}
\usepackage{mathpartir}
\usepackage{xcolor}
\usepackage{amsmath}
\usepackage{cleveref}
\usepackage{subcaption}
\usepackage{todonotes}
\usepackage{csquotes}

\usepackage{listings}
\usepackage{smalite}
\usepackage{biblatex}
\usepackage{mathtools}
\usepackage{syntax}

\newlength{\mygl}

\newcommand{\spluss}{\kern.01em\raisebox{.92ex}{\scalebox{.6}{+}}}
\newcommand{\splus}{\spluss\kern.2em}
\newcommand{\sstars}{\kern.01em\raisebox{.92ex}{\scalebox{.6}{*}}}
\newcommand{\sstar}{\sstars\kern.2em}
\newcommand{\sqmark}{\kern.01em\raisebox{.92ex}{\scalebox{.6}{?}}\kern.2em}
\newcommand{\altl}{\ |\ }

\usepackage{tikz}
\usetikzlibrary{
  arrows,
  chains,
  decorations,
  decorations.pathreplacing,
  decorations.text,
  fit,
  shadows.blur,
  shapes
}

\NewDocumentCommand\tikzbox{t+mm}{
  \path (-#2, -#3) coordinate (bottom left)
        ( #2,  #3) coordinate (top right);
  \path (bottom left |- top right) coordinate (top left);
  \path (bottom left -| top right) coordinate (bottom right);
  \path[use as bounding box] (bottom left) rectangle (top right);

  \IfBooleanT{#1}{
    \draw[help lines,thick,gray!15,step=.5]
    (bottom left) grid (top right) (0.0, 0.0) [fill]circle (5pt);
  }
}

\newcommand{\xvdashsmall}[1]{\mkern-7mu\raisebox{.25ex}{\scalebox{.45}{\textrm{#1}}}}
\newcommand{\xvdash}[2]{%
  \vdash_{\xvdashsmall{#1}}^{\xvdashsmall{#2}}%
}

\usepackage{myacronym}
\newacro{AST}{Abstract Syntax Tree}
\newacro{GUI}{Graphical User Interface}
\newacro{UIDL}{User Interface Description Language}
\newacro{HMI}{Human-Machine Interface}
\newacro{IR}{Intermediate Representation}
\newacro{RSSA}{Reactive Static Single Assignment}
\newacro{TCAS}{Traffic Alert and Collision Avoidance System}
\newacro{WP}{Weakest Precondition}

\newcommand{\set}[1]{\{ #1 \}}
\newcommand{\vroot}{\var{r}}

\newcommand{\rulename}[1]{#1}

\newcommand{\smala}{Smala}
\newcommand{\smalite}{Smalite}
\newcommand{\velus}{Vélus}

\newcommand{\decljudgement}[1]{\hfill\framebox{\mbox{\ensuremath{#1}}}}

\title{Reactive Semantics for User Interface Description Languages}

\author{Basile Pesin \email{basile.pesin@enac.fr}
  \and Celia Picard \email{celia.picard@enac.fr}
  \and Cyril Allignol \email{cyril.allignol@enac.fr}
  \institute{Fédération ENAC ISAE-SUPAERO ONERA, Université de Toulouse, France}
}

\date{June 20, 2025}
\newcommand{\titlerunning}{Reactive Semantics for User Interface Description Languages}
\newcommand{\authorrunning}{B. Pesin, C. Picard, C. Allignol}
\newcommand{\event}{ICE 2025}

\begin{document}

\maketitle

\begin{abstract}
  User Interface Description Languages (UIDLs) are high-level languages that facilitate the development of Human-Machine Interfaces, such as Graphical User Interface (GUI) applications.
They usually provide first-class primitives to specify how the program reacts to an external event (user input, network message), and how data flows through the program.
Although these domain-specific languages are now widely used to implement safety-critical GUIs, little work has been invested in their formalization and verification.

In this paper, we propose a denotational semantic model for a core reactive UIDL, Smalite, which we argue is expressive enough to encode constructs from more realistic languages.
This preliminary work may be used as a stepping stone to produce a formally verified compiler for UIDLs.

\end{abstract}

\section{Introduction and Context}

With the democratization of interactive devices, the general interaction paradigm has changed.
Users now expect to interact with their systems through tactile interactions or advanced interactive devices.
This extends to critical systems too, notably aviation related ones.
For instance, aircraft cockpits are now digital and tactile~\cite{LIM20181} and paper strips for air traffic controllers have been replaced by elaborate digital dashboards~\cite{huber2020}.

Using traditional programming languages to implement the interactive parts of systems implies the use of numerous callbacks.
The resulting code quickly becomes spaghetti code and gets particularly difficult to analyze and maintain~\cite{maier2010deprecating,martin2022causette}.
The use of dedicated languages has shown to be beneficial in such circumstances~\cite{myers1991separating}. Such languages are called \acp{UIDL}.
These languages can efficiently describe both the appearance of the system (its scene graph) and the interactive behavior (mainly activation propagation).
They are becoming more popular, including for the development of critical systems~\cite{Conversy:Volta:2017,anand2022avionics}.
However, these languages have not been formally defined, particularly not their semantics, which hinders any formal reasoning on these safety-critical programs.
In this paper, we tackle this problem by proposing denotational semantics for a minimal declarative \ac{UIDL}.

There has been a few other attempts to give a formal semantics to interactive languages or libraries such as React~\cite{Madsen:EssenceOfReact:2020}.
However, React is based on a functional language and as such, has very different mechanisms than other \ac{UIDL}/interactive languages.
Indeed, the behavior of a React program is described using small-step operational
semantics which specify the order in which components are rendered.
In this paper, we focus on a declarative language with denotational semantics
where the order of updates is left implicit.

Previous work has paved the way towards a minimal common abstract syntax for Smala, a declarative \ac{UIDL}~\cite{Magnaudet:Smala:2018}.
A first formal semantic model based on bigraphs has been proposed~\cite{Nalpon:UIDL:2022}, but in this work we explore an alternative expression of the semantics, which is simpler and should facilitate compiler verification in the fashion of CompCert~\cite{Leroy:Realistic:2009} or \velus{}~\cite{Bourke:Velus:2023}. To give more confidence in those semantics, we also aim at proving the equivalence between the two semantics, as future work.

Hence, we propose a new denotational semantics for Smalite, a declarative
\ac{UIDL} which includes a minimal set of constructs from Smala.
As a preliminary step to a compiler correctness proof, we mechanized the
language and its semantics in the Rocq prover \footnote{formerly known as
Coq}~\cite{Coq:2024}, and implemented a prototype compiler that generates
imperative code.
In \cref{sec:example}, we give an informal presentation of our language through
an example program implementing a simple \ac{GUI} program.
Then, in \cref{sec:semantics}, we describe our formalization of the reactive semantics
of the language.
Finally, in \cref{sec:future}, we discuss the future steps towards extending the
language and building a formal proof of correctness for its compiler.

\section{Specifying interactions with \smala{}}
\label{sec:example}

\begin{figure}[!ht]
  \begin{tikzpicture}
    \tikzbox{\textwidth/2}{8}
    \node[anchor=west,text width=\textwidth] at (-\textwidth/2,0) {
\begin{lstlisting}[numbers=left,language=smalite]
Component root {
  Int count 3;

  Spike zero; (count == 0) -> zero;

  Frame f ("ICE 2025", 300, 50) {
    Font _ ("arial.ttf", 20) {
      FillColor btn1 (150,150,150) {
        Rectangle r (0,0,100,f.height) {
          FillColor _ (0,0,0) {
            Text _ ("decr", r.x + 30, 13) {}
          };
        };
        r.press -> hg; hg: 255 =: green;
        r.release -> dhg; dhg: 150 =: green;
      };
      btn1.r.release -> dec; dec: last count-1 =: count;
      zero ->! btn1.r; (count > 0) -> btn1.r;

      FillColor btn2 (150,150,150) {
        Rectangle<d> r (btn1.r.width+10,0,100,f.height) {
          FillColor _ (0,0,0) {
            Text t ("restart", r.x + 20, 13) {}
          };
        };
        r.press -> hg; hg: 255 =: green;
        r.release -> dhg; dhg: 150 =: green;
      };
      btn2.r.release -> rst; rst: 3 =: count;
      (count < 3) -> btn2.r; (count == 3) ->! btn2.r;

      FillColor _ (255,255,255) {
        Text t ("rem: " + str(count), btn2.r.x+btn2.r.width+10, 13) {
          count -> chtext; chtext: "rem: " + str(count) =: text;
        };
      }
    }
  };

  Exit e (0) { f.close -> trigger };
}
\end{lstlisting}
    };

    \begin{scope}[yshift=-1cm,xshift=1cm]
    \node (ex3) at (4.2,7.5) {\includegraphics[width=5cm]{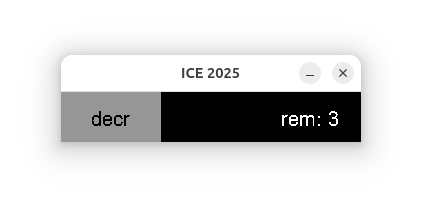}};
    \node (ex3hg) at (4.2,5) {\includegraphics[width=5cm]{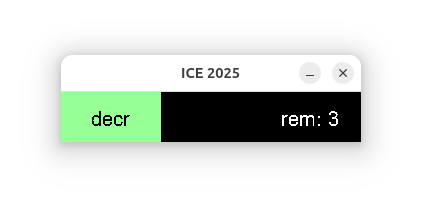}};
    \node (ex2) at (4.2,2.5) {\includegraphics[width=5cm]{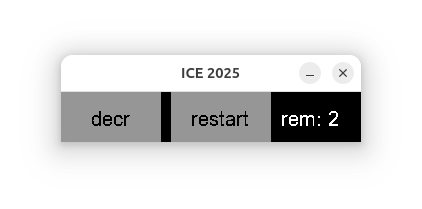}};
    \node (ex1) at (4.2,0) {\includegraphics[width=5cm]{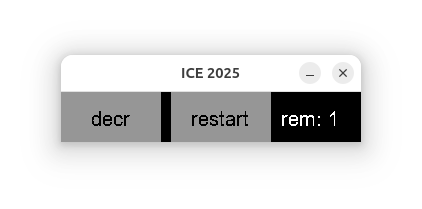}};
    \node (ex0) at (4.2,-2.5) {\includegraphics[width=5cm]{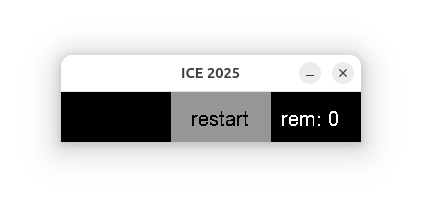}};

    \draw[-latex] ($(ex3.south) + (0,.8)$) -- ($(ex3hg.north) - (0,.65)$)
    node[left,midway] {press \texttt{decr}};
    \draw[-latex] ($(ex3hg.south) + (0,.8)$) -- ($(ex2.north) - (0,.65)$)
    node[left,midway] {release \texttt{decr}};
    \draw[-latex,dashed] ($(ex2.south) + (0,.8)$) -- ($(ex1.north) - (0,.65)$)
    node[left,midway] {\texttt{decr}};
    \draw[-latex,dashed] ($(ex1.south) + (0,.8)$) -- ($(ex0.north) - (0,.65)$)
    node[left,midway] {\texttt{decr}};
    \end{scope}

    \draw[-latex,dashed]
    ($(ex2.east) - (.8,0)$) to[bend right=20] ($(ex3.east) - (.8,0)$);
    \draw[-latex,dashed]
    ($(ex1.east) - (.8,0)$) to[bend right=20] ($(ex3.east) - (.8,0)$);
    \draw[-latex,dashed]
    ($(ex0.east) - (.8,0)$) to[bend right=20]
    node[pos=.891,left] {\texttt{restart}} ($(ex3.east) - (.8,0)$);

  \end{tikzpicture}
\caption{An example Smala program}
\label{fig:example}
\end{figure}
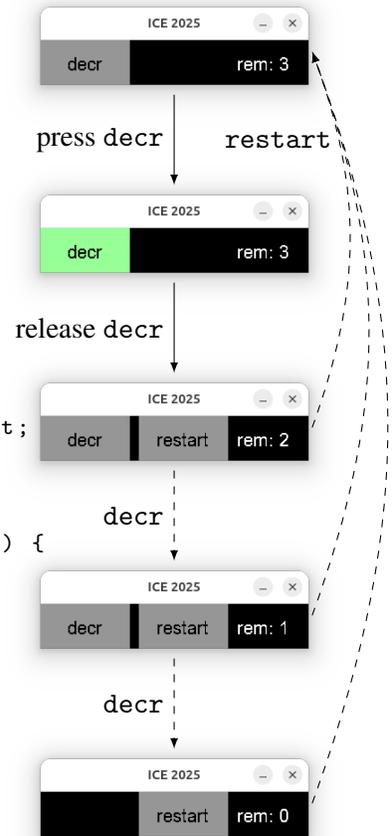

The language we propose closely resembles \smala{}~\cite{Magnaudet:Smala:2018}, a
\ac{UIDL} used in safety-critical applications~\cite{Conversy:Volta:2017, cousy1}.
In \cref{fig:example}, we present an example \smala{} program where two buttons
control a counter by either decrementing it until it reaches 0 or setting it
back to 3.

\smala{} programs are composed of named \emph{processes}.
The root process is a \kw{Component} (\texttt{line 1}), usually named \texttt{root}, which
contains child processes.
The counter is implemented as a \emph{property} \texttt{count}
(\texttt{line 2}), which is declared with a type and an expression giving its
initial value.
Then, the program contains a \kw{Spike} (\texttt{line 3}), which represents an
event that may be triggered from inside or outside the program, and reacted to.
This particular \kw{Spike} represents the event ``the counter has just reached
0''.
Indeed, it is triggered by the \emph{binding} on the same line, whose
left-hand side is a condition checking that \texttt{count} equals 0.
Conditions are checked only at reactions where the value of one of the
properties involved changes; therefore, \texttt{zero} is triggered only at
reactions where the value of \texttt{count} changes to 0.

The next process is a \kw{Frame}, which is a special
\emph{graphical} component (\texttt{line 6}).
For the purpose of our reactive semantics, a \kw{Frame} is just syntactic sugar
for a component with properties \texttt{title}, \texttt{width} and
\texttt{height}, passed as parameters between parentheses, as well as a
\kw{Spike} \texttt{close}.
On the other hand, our prototype compiler generates code which uses the
\kw{Frame} and its parameters to open a window.
The compiler also binds user requests to close the window (e.g. by clicks on the
\begin{tikzpicture}[baseline = -0.7ex]
  \draw[lightgray!50, fill = lightgray!50] (0, 0) circle (1ex);
  \draw[gray, very thick] (-0.5ex, 0.5ex) -- (0.5ex, -0.5ex);
  \draw[gray, very thick] (-0.5ex, -0.5ex) -- (0.5ex, 0.5ex);
\end{tikzpicture}
button) to triggering the \texttt{close} \kw{Spike}.
Therefore, the program can define custom reactions to user actions.

The first child of the \kw{Frame} defines the \kw{Font} used to display text in
the interface (\texttt{line 7}).
Since the rest of the program does not need to refer to this particular
component, its name is unspecified (\texttt{_}).
Then, the frame contains three main components: two buttons, and a text label.
Each button is implemented by a component \kw{FillColor} (with parameters
\texttt{red}, \texttt{green} and \texttt{blue}), which sets the color of its
\kw{Rectangle} child (with parameters \texttt{x}, \texttt{y}, \texttt{width} and
\texttt{height}).
Text labels are implemented by a \kw{FillColor} which sets the color of its
\kw{Text} child (with parameters \texttt{text}, \texttt{x} and \texttt{y}).

\kw{Rectangle} components have two predefined spikes, \texttt{press}
and \texttt{release}, which are respectively triggered when the mouse is
pressed/released while the mouse pointer is on top of the rectangle.
The logic of each button is implemented as a reaction to these spikes: on
\texttt{press}, the button is highlighted by increasing the \texttt{green}
property of its \kw{FillColor} (\texttt{lines 14, 26}).
On \texttt{release}, the \texttt{green} component is set back to its default
value (\texttt{lines 15, 27}).
The graphical effect of these first two reactions are shown at right of
\cref{fig:example}.

The \texttt{release} \kw{Spike} of each button is bound to an action on the counter.
Releasing \texttt{decr} decreases the counter (\texttt{line 17}) by setting its new value to its previous value (accessed with \kw{last}) minus one.
Releasing \texttt{restart} sets the counter back to 3 (\texttt{line 29}).
Buttons may be activated or deactivated depending on the value of the counter.
If the counter equals 0, it is not possible anymore to decrease it, and
therefore the \texttt{decr} button is deactivated (first binding \hbox{\texttt{->! btn1.r}}
on \texttt{line 18}).
As soon as the counter goes back above 0, it is reactivated.
Conversely, the \texttt{restart} button is deactivated when the counter equals
3, and reactivated otherwise (\texttt{line 30}).
Normally, the activation of a parent automatically activates all of its
children.
However, since the counter starts at 3, the \texttt{restart} button must be initially
deactivated, which is specified by \smainit{d} on \texttt{line 21}.

The last child component of the \kw{Frame} is a \kw{Text} label
displaying the value of \texttt{count}.
It is updated every time the value of \texttt{count} changes
(binding \texttt{count ->} on \texttt{line 34}).

Finally, the program contains an \kw{Exit} component, which takes an exit
code as a parameter, and exposes a \kw{Spike} \texttt{trigger} which halts the
program.
The \texttt{close} \kw{Spike} of the \kw{Frame} is bound to \texttt{trigger},
which allows the user to close the program by closing the window.

\section{Formalizing \smalite{}'s semantics}
\label{sec:semantics}

In this paper, we define the formal semantics of \smalite{}, a more
restricted form of \smala{}.
\smalite{} is a reactive language: the execution of a program can be seen as a
series of reactions to external events.
We saw in the example of \cref{fig:example} that the main type of external event is the triggering of
a \kw{Spike}, such as \texttt{close} (\texttt{line 40}) or \texttt{release} (\texttt{line 15}).
Another possible external event not showcased in the example would be an outside
modification of a property.
For instance, an adaptive \ac{GUI} program could listen for the
resizing of a \kw{Frame} which would be encoded as a change of its
\texttt{width} or \texttt{height} property.

A reaction may then update the state of the process by \emph{activating} or
\emph{deactivating} permanent processes, or \emph{assigning} values to
properties.
A reaction may also \emph{trigger} spikes which, as seen in the
example, may have an effect on the view.
Therefore, we generalize external events to also account for events triggered by
the program during a reaction, as defined below.
\begin{align*}
\var{ev} ::=\ \etrigger{\var{path}}
\altl \eassign{\var{v}}{\var{path}}
\altl \eactivate{\var{path}}
\altl \edeactivate{\var{path}}
\end{align*}

First, an event can be the triggering of a \enquote{transient} process, which does not
retain its activation between cycles (\kw{Spike} or assign).
Second, it can be an assignment to a property.
Last, it can be the activation or deactivation of a \enquote{permanent} process, which
retains its activation between cycles (\kw{Component} or binding).

\subsection{Abstract syntax of \smalite{}}

\begin{figure}
  \setlength{\jot}{0pt}
  \begin{minipage}{.35\textwidth}
\begin{align*}
  p ::=&\ \smaprop{\var{ty}}{x}{e} \\
  \altl& \smaspike{x} \\
  \altl& \smaassign{x}{e}{\var{path}} \\
  \altl& \smabinding{x}{\var{lhs}}{\var{ia}}{\var{rhs}} \\
  \altl& \smacomponent{\var{ia}}{x}{p\sstar}
\end{align*}
\end{minipage}
  \hfill
  \begin{minipage}{.62\textwidth}
\begin{align*}
  \var{path} ::=&\ \var{x} \altl \var{x} \texttt{.} \var{path}\\
  e ::=&\ c
  \altl \var{path}
  \altl \kw{last}\ \var{path}
  \altl \diamond e
  \altl e \oplus e\\
  \var{ia} ::=&\ \texttt{a} \altl \texttt{d}\\
  \var{lhs} ::=&\ \lhstrigger{\var{path}}
  \altl \lhsactivate{\var{path}}
  \altl \lhsdeactivate{\var{path}}
  \altl \lhschange{\var{path}}
  \altl \lhscond{e}\\
  \var{rhs} ::=&\ \rhstrigger{\var{path}}
  \altl \rhsactivate{\var{path}}
  \altl \rhsdeactivate{\var{path}}
\end{align*}
\end{minipage}
\caption{Abstract Syntax of Smalite}
\label{fig:syntax}
\end{figure}

\begin{figure}
  \begin{center}
    \small
  \begin{tabular}{c|c||c|c||c}
    Concrete LHS & Concrete RHS & Abstract LHS & Abstract RHS & Event \\
    \hline
    \texttt{path ->} & \texttt{-> path} & \lhstrigger{\var{path}} & \rhstrigger{\var{path}} & \etrigger{\var{path}} \\
    \texttt{path ->} & N/A & \lhschange{\var{path}} & N/A & \eassign{\var{v}}{\var{path}} \\
    \texttt{(e) ->} & N/A & \lhscond{\var{e}} & N/A &
    \eassign{\var{v}}{\var{path}}, \var{path} $\in{}$ \textsf{free}(\var{e}) \\
    \texttt{path ->} & \texttt{-> path} & \lhsactivate{\var{path}} & \rhsactivate{\var{path}} & \eactivate{\var{path}} \\
    \texttt{path !->} & \texttt{->! path} & \lhsdeactivate{\var{path}} & \rhsdeactivate{\var{path}} & \edeactivate{\var{path}} \\
  \end{tabular}
  \end{center}
  \caption{Correspondence between concrete syntax, abstract syntax and events}
  \label{fig:event-corres}
\end{figure}

We now detail the abstract syntax of \smalite{} as we formalize it.
A minimalized version of \smala{} has been proposed in previous work~\cite{Magnaudet:Smala:2018}.
It includes a small set of core elements that constitutes the heart of the language,
expressive enough to describe the full set of components of the whole language while
minimizing the number of constructs to actually formally define and verify.
The translation from \smala{} to
this minimal set is done through a transpilation pass not described here.
Our syntax is presented in \cref{fig:syntax}, and builds upon the one proposed
previously~\cite{Magnaudet:Smala:2018}.

A program is a process \var{p}.
There are five different processes detailed hereafter.
A process may be the declaration of a
property with its type \var{ty}, name \var{x} and initial value \var{e}.
There are two kinds of transient processes, which are triggered during a
single cycle: spikes and assignments.
The latter assigns the result of the evaluation of an expression \var{e} to a
property specified by an absolute \var{path}.
An expression is either a constant \var{c}, the \var{path} to access the value
or the \kw{last} value of a property, or a unary ($\diamond$) or binary ($\oplus$)
operation.
Types, constants and operations are those of the host language that the compiler
targets (in the case of our prototype compiler, CompCert C).

Finally, there are two kinds of permanent processes, which retain their
activation between cycles.
Both kinds may be initially activated or deactivated following the \var{ia}
flag.
The first permanent process is the binding, denoted by an arrow \texttt{->}.
It binds the event specified by its left-hand-side to
the event specified by its right-hand-side.
An event detected by a left-hand-side may be the triggering (\texttt{T?}) of a transient
process (\kw{Spike}, assign), the activation (\texttt{A?}) or deactivation (\texttt{D?}) of a permanent
process (binding, \kw{Component}), the change of value of a property
(\texttt{C?}), or a condition being true after one of its free variable changes
(\texttt{(\var{e})?}).
Right-hand-sides may trigger a transient process (\texttt{T!}) or activate (\texttt{A!}) or
deactivate (\texttt{D!}) a permanent process.
The second permanent process is the component, which contains a list of
sub-processes.

This abstract syntax is sligtly different from the concrete syntax used for the example.
First, each process is explicitly named.
Second, there is only one generic \kw{Component}. It can then be instantiated to
create all other components,
particularly the graphical components (\kw{Frame}, \kw{Rectangle}, etc), as
those do not play a special role in the semantics.
Third, as we have seen, the left- and right-hand sides of bindings distinguish
between triggering of transient
processes, activation/deactivation of permanent processes, and change of property value.
\Cref{fig:event-corres} recapitulates the correspondence between concrete left- and
right-hand-sides, as seen in the example, and their abstract counterpart.
Last, processes refer to each other using their absolute paths, which are made up of
a sequence of identifiers indicating the path from the root process to the
process of interest.

In practice, our prototype compiler parses the source program of
\cref{fig:example} into an \ac{AST} represented as an inductive type in Rocq.
Then, a sequence of functions recursively elaborates it into a second \ac{AST}
representing the more restricted syntax of \cref{fig:syntax}, in three passes.
First, the elaborator fills-in missing names by generating globally unique
identifiers.
Then, it expands graphical components into generic \kw{Component} by adding
the predefined properties and spikes (e.g. \texttt{title}, \texttt{width},
\texttt{frame} for a \kw{Frame}).
Finally, it type-checks the program, makes relative paths absolutes, differentiates
transient and permanent processes in bindings, and builds the
restricted \ac{AST}.

\subsection{Semantics of program initialization}

\begin{figure}[!t]
  \begin{subfigure}{\textwidth}
    \decljudgement{\xvdash{init}{activ} \var{p}_{\vroot} \Downarrow A}
  \begin{mathpar}
    \inferrule* [Right=\rulename{IAP}]
        { }
        {\xvdash{init}{activ} (\smaprop{\var{ty}}{x}{e})_{\vroot_x} \Downarrow \emptyset}
    \and
    \inferrule* [Right=\rulename{IAS}]
        { }
        {\xvdash{init}{activ} (\smaspike{x})_{\vroot_x} \Downarrow \emptyset}
        \and
    \inferrule* [Right=\rulename{IAA}]
        { }
        {\xvdash{init}{activ} (\smaassign{x}{e}{p})_{\vroot_x} \Downarrow \emptyset}
    \\
    \inferrule* [Right=\rulename{IAB$_1$}]
       {  }
       { \xvdash{init}{activ} (\smabinding{x}{\var{lhs}}{d}{\var{rhs}})_{\vroot_x} \Downarrow \emptyset }
    \and
   \inferrule* [Right=\rulename{IAB$_2$}]
       {  }
       { \xvdash{init}{activ} (\smabinding{x}{\var{lhs}}{a}{\var{rhs}})_{\vroot_x} \Downarrow \set{\vroot_x.x}}
    \\
    \inferrule* [Right=\rulename{IAC$_1$}]
       {  }
       { \xvdash{init}{activ} (\smacomponent{d}{x}{\var{ps}})_{\vroot_x} \Downarrow \emptyset }
    \and
   \inferrule* [Right=\rulename{IAC$_2$}]
       { \forall i.\ \xvdash{init}{activ} (\var{ps}_i)_{\vroot_x.x} \Downarrow A_i }
       { \xvdash{init}{activ} (\smacomponent{a}{x}{\var{ps}})_{\vroot_x} \Downarrow (\bigcup\nolimits_i A_i) \cup \set{\vroot_x.x} }
  \end{mathpar}
  \subcaption{Initialization of activation}
  \label{fig:rules-init-activ}
  \end{subfigure}
  \begin{subfigure}{\textwidth}
    \decljudgement{E \xvdash{init}{env} \var{p}_\vroot{} \Downarrow E'}
    \begin{mathpar}
    \inferrule* [Right=\rulename{IES}]
        { }
        {E \xvdash{init}{env} (\smaspike{x})_{\vroot_x} \Downarrow E}
        \and
    \inferrule* [Right=\rulename{IEA}]
        { }
        {E \xvdash{init}{env} (\smaassign{x}{e}{p})_{\vroot_x} \Downarrow E}
    \and
    \inferrule* [Right=\rulename{IEB}]
       {  }
       {E \xvdash{init}{env} (\smabinding{x}{\var{lhs}}{_}{\var{rhs}})_{\vroot_x} \Downarrow E}
    \and
    \inferrule* [Right=\rulename{IEP}]
        { \emptyset, E \xvdash{exp}{} e \Downarrow v }
        {E \xvdash{init}{env} (\smaprop{\var{ty}}{x}{e})_{\vroot_x} \Downarrow E[\vroot_x.x \mapsto v]}
        \and
   \inferrule* [Right=\rulename{IEC$_1$}]
       {  }
       { E \xvdash{init}{env} (\smacomponent{_}{x}{\epsilon})_{\vroot_x} \Downarrow E }
        \and
    \inferrule* [Right=\rulename{IEC$_2$}]
       { E \xvdash{init}{env} p_{(\vroot_x.x)} \Downarrow E' \\
         E' \xvdash{init}{env} (\smacomponent{_}{x}{\var{ps}})_{\vroot_x} \Downarrow E'' }
       { E \xvdash{init}{env} (\smacomponent{_}{x}{p; \var{ps}})_{\vroot_x} \Downarrow E'' }
       \end{mathpar}
    \subcaption{Initialization of environment}
    \label{fig:rules-init-env}
    \end{subfigure}
  \begin{subfigure}{\textwidth}
   \decljudgement{\xvdash{init}{} \var{p} \Downarrow E, A}
    \begin{mathpar}
    \inferrule* [Right=\rulename{I}]
        {\xvdash{init}{activ} \var{p}_{\epsilon} \Downarrow A \\
         \emptyset \xvdash{init}{env} \var{p}_{\epsilon} \Downarrow E }
        {\xvdash{init}{} \var{p} \Downarrow E, A }
  \end{mathpar}
    \subcaption{Program initialization}
    \label{fig:rules-init-prog}
  \end{subfigure}
\caption{Initialization rules}
\label{fig:rules-init}
\end{figure}

We now describe the rules that specify the initialization of the system, presented in \cref{fig:rules-init}.
The first judgment,
$\xvdash{init}{activ} \var{p}_{\vroot} \Downarrow A$, specifies the
initial activation for a process \var{p} rooted at path
\vroot{}, where \var{A} is a set of paths of permanent processes.
It is defined by the rules in \cref{fig:rules-init-activ}.
Non-permanent processes (property, spike, assignment) never appear in this set.
Bindings and components appear in the set if-and-only-if they are declared with
\smainit{a}.
For a component marked with \smainit{a}, the sub-processes are initially
activated according to the same rules.

\begin{figure}
  \decljudgement{E, E' \xvdash{exp}{} e \Downarrow v}
  \begin{mathpar}
    \inferrule* [Right=\rulename{EC}]
       {  }
       { E, E' \xvdash{exp}{} c \Downarrow c^{\#} }
    \and
    \inferrule* [Right=\rulename{EV}]
       {  }
       { E, E' \xvdash{exp}{} x \Downarrow E'(x) }
    \and
    \inferrule* [Right=\rulename{EL}]
       {  }
       { E, E' \xvdash{exp}{} \kw{last}\ x \Downarrow E(x) }
    \and
    \inferrule* [Right=\rulename{EU}]
       { E, E' \xvdash{exp}{} e_1 \Downarrow v_1 \\
         \diamond^{\#}(v_1) = \lfloor v \rfloor }
       { E, E' \xvdash{exp}{} \diamond\ e_1 \Downarrow v }
     \and
     \inferrule* [Right=\rulename{EB}]
       { E, E' \xvdash{exp}{} e_1 \Downarrow v_1 \\
         E, E' \xvdash{exp}{} e_2 \Downarrow v_2 \\
         \oplus^{\#}(v_1, v_2) = \lfloor v \rfloor }
       { E, E' \xvdash{exp}{} e_1\ \oplus e_2 \Downarrow v }
  \end{mathpar}
\caption{Expression evaluation rules}
\label{fig:rules-exp}
\end{figure}

The second judgment,
$E \xvdash{init}{env} \var{p}_{\vroot} \Downarrow E'$, specifies how the
initial property values are added to initial environment $E$ to form $E'$, where
environments are represented by finite maps from property paths to values.
It is defined by the rules in \cref{fig:rules-init-env}.
As expected, spikes, assignments and bindings have no effect on the initial environment.
For properties (rule \rulename{IEP}), the initialization expression is evaluated under the starting environment $E$ and the resulting value is added to $E$.
A value is either a numerical value from the C language (integer, floating-point number), a boolean value, or a string of characters.
The rules for expression evaluation specifying
$E, E' \xvdash{exp}{} e \Downarrow v$ are unsurprising, and presented in
\cref{fig:rules-exp}.
The notation $c^{\#}$ represents the semantic denotation of constant $c$
(respectively $\diamond^{\#}$ for unary operator $\diamond$ and $\oplus^{\#}$ for binary operator $\oplus$).
The current value of a variable is searched in $E'$, the updated
environment (rule \rulename{EV}), while its \kw{last} value is searched in $E$, which represents the
environment at the end of the last step, and is empty at initialization (rule \rulename{EL}).
The evaluation of operators may fail (e.g. in case of a division by zero), which
is denoted on their return value by $\lfloor v \rfloor$.

Rules \rulename{IEC$_1$} and \rulename{IEC$_2$}, for components, highlight an interesting design choice.
In our definitions, the environment is updated sequentially, following the order the sub-processes are written in the program.
One simpler definition would have been to update the same environment $E$ in parallel for each sub-process $\var{ps}_i$ into an environment $E_i$, and to take the union of these environments, but this would have prevented writing initial expressions that depend on the values of other properties, as seen for the button text position in \cref{fig:example} (\texttt{line 21}).
Conversely, a more expressive semantics would allow the initialization of a property to depend on the values of properties later in the program, as long as there is no cycle in the definitions.
This would however require (1)~a more complicated semantic judgment for initialization, (2)~an analysis of the absence of cycle at the source level, and (3)~a way to schedule property initialization in the compiled code.
Although these are all feasible, we believe that the marginal expressivity gains are not sufficient to justify this additional complexity.

Finally, the two judgments described above are combined in rule \rulename{I}, presented in \cref{fig:rules-init-prog}.
It specifies the initialization of a \smalite{} program: $\xvdash{init}{} \var{p} \Downarrow E, A$ indicates that the process \var{p}, rooted at the empty path $\epsilon$, is initialized with initial environment $E$ and initial activation $A$.

\subsection{Semantics of reaction}
To describe how a program reacts to an external event, let us start with the
main reaction rule presented in \cref{fig:rule-react}.
The judgment
$E, A \xvdash{react}{} \reactevent{\var{p}}{\var{ev}} \Downarrow E', A', T'$
indicates that, with initial environment $E$ and activation $A$, the process
\var{p} reacts to an external event \var{ev} by updating its environment
to $E'$, its activation to $A'$, and emitting a set $T$ of external events.
The relation between these parameters are specified by three premises, which we
now detail.

\begin{figure}
  \decljudgement{E, A \xvdash{react}{} p(\var{ev}) \Downarrow E', A', T }
\begin{mathpar}
  \inferrule* [right=R]
      { T = \set{\var{ev}_0} \cup \set{\var{ev}\ |\ E, A,
        T, E', A' \xvdash{prop}{proc} \var{p} \Downarrow \var{ev}} \\
        E, T, E', A' \xvdash{safe}{proc} \var{p}_{\epsilon} \\
        E, A \xvdash{update}{} T \Downarrow E', A' }
      { E, A \xvdash{react}{} \reactevent{\var{p}}{\var{ev}_0} \Downarrow E', A',
        (T \cap \set{ \etrigger{path}\ |\ \smaspike{path} \in \var{p}}) }
\end{mathpar}
\caption{Main reaction rule}
\label{fig:rule-react}
\end{figure}

\subsubsection{Event propagation}
The first premise determines the set $T$ of events that are emitted
during the reaction step.
$T$ contains the external event that triggered the reaction, \var{ev}.
It also contains any additional event propagated by the process: this is represented by
judgment
\hbox{$E, A, T, E', A' \xvdash{prop}{proc} \var{p} \Downarrow \var{ev}$}
which can be read as
\enquote{when process \var{p} reacts to events $T$, it also emits event \var{ev}}.
The rules defining this judgment are presented in \cref{fig:rules-prop-1,fig:rules-prop-2}.
These rules are not syntax-directed: instead, each one asserts the
existence of a sub-process that may propagate events.
We write the judgement $\smafindproc{p}{\var{path}}{p'}$ for \enquote{there is a process \var{p'} rooted at \var{path} in \var{p}}.

\begin{figure}[t]
    \decljudgement{E, T, E' \xvdash{prop}{lhs} \var{lhs}}
    \begin{mathpar}
    \inferrule* [Right=\rulename{PLT}]
        { \etrigger{x} \in T }
        { E, T, E' \xvdash{prop}{lhs} \lhstrigger{x} }
    \and
    \inferrule* [Right=\rulename{PLA}]
        { \eactivate{x} \in T }
        { E, T, E' \xvdash{prop}{lhs} \lhsactivate{x} }
    \and
    \inferrule* [Right=\rulename{PLD}]
        { \edeactivate{x} \in T }
        { E, T, E' \xvdash{prop}{lhs} \lhsdeactivate{x} }
     \and
     \inferrule* [Right=\rulename{PLC}]
        { \eassign{v}{x} \in T }
        { E, T, E' \xvdash{prop}{lhs} \lhschange{x} }
     \and
     \inferrule* [Right=\rulename{PLI}]
        { x \in \mathsf{free}(e) \\
          \eassign{v}{x} \in T \\
          E, E' \xvdash{exp}{} e \Downarrow \typ{true} }
        { E, T, E' \xvdash{prop}{lhs} \lhscond{e} }
    \end{mathpar}
    \decljudgement{A, A' \xvdash{prop}{rhs} \var{rhs} \Downarrow \var{ev}}
    \begin{mathpar}
     \inferrule* [Right=\rulename{PRT}]
       { \vroot_x \in A' }
       { A, A' \xvdash{prop}{rhs} \rhstrigger{\vroot_x.x} \Downarrow \etrigger{\vroot_x.x} }
     \and
     \inferrule* [Right=\rulename{PRA}]
       { \vroot_x \in A' \\ \vroot_x.x \notin A }
       { A, A' \xvdash{prop}{rhs} \rhsactivate{\vroot_x.x} \Downarrow \eactivate{\vroot_x.x} }
     \and
     \inferrule* [Right=\rulename{PRD}]
       { \vroot_x \in A' \\ \vroot_x.x \in A }
       { A, A' \xvdash{prop}{rhs} \rhsdeactivate{\vroot_x.x} \Downarrow
         \edeactivate{\vroot_x.x} }
    \end{mathpar}
    \decljudgement{E, A, T, E' A' \xvdash{prop}{proc} \var{p} \Downarrow \var{ev}}
    \begin{mathpar}
    \inferrule* [Right=\rulename{PB}]
        { \smafindproc{\var{p}}{\vroot_x.x}{\smabinding{x}{\var{lhs}}{_}{\var{rhs}}} \\
          \vroot_x.x \in A' \\
          E, T, E' \xvdash{prop}{lhs} \var{lhs} \\
          A, A' \xvdash{prop}{rhs} \var{rhs} \Downarrow \var{ev} \\
        }
        { E, A, T, E', A' \xvdash{prop}{proc} \var{p} \Downarrow \var{ev} }
        \and
    \inferrule* [Right=\rulename{PA}]
        { \smafindproc{\var{p}}{\vroot_x.x}{\smaassign{x}{e}{\vroot_y.y}} \\
          \vroot_x \in A' \\
          \vroot_y \in A' \\
          \etrigger{\vroot_x.x} \in T\\
          E, E' \xvdash{exp}{} e
          \Downarrow v \\
        }
        { E, A, T, E', A' \xvdash{prop}{proc} \var{p} \Downarrow
          \eassign{v}{\vroot_y.y}  }
  \end{mathpar}
  \caption{Propagation rule for bindings and assignments}
  \label{fig:rules-prop-1}
  \end{figure}

The first set of rules, in \cref{fig:rules-prop-1}, describes how bindings propagate events.
Rule \rulename{PB}, presented at the bottom, specifies that a binding only
propagates events if it is activated.
The third premise $E, T, E' \xvdash{activ}{lhs} \var{lhs}$ of the rule assumes that the left-hand side is activated by a matching event.
In particular, if the left-hand side is a condition, it must evaluate to \typ{true}.
The final premise $A, A' \xvdash{prop}{rhs} \var{rhs} \Downarrow \var{ev}$ specifies which event \var{ev} the right-hand side emits.
All rules specifying this judgment mandate that the parent of the process involved in the event is active.
Activation (resp. deactivation) events are only emitted if the process was
previously deactivated (resp. activated): in other words, ``non-events'' that do
not modify the activation state are not propagated.

The last rule in \cref{fig:rules-prop-1}, covers the assignment.
It states that an assignment produces an \eassign{v}{y} event
when (1)~its parent is activated, (2)~the parent of $y$ is activated, (3)~the
assignment has been triggered by another event, and (4) the expression of the
assignment evaluates to value $v$.
Here, we made one interesting choice:
the second premise implies that it is not possible to assign to a
property whose parent is deactivated.
This constraint was added because
the behavior of programs where a property is assigned to when its parent
is deactivated was not clear: should the assignment event be propagated during
the reaction, or when the parent becomes active again, or not at all ?
We chose to eliminate this question by forbidding this case entirely, as we believe it is not useful in real-world programs.
In practice, our prototype compiler statically checks that this situation can
never arise.

  \begin{figure}[t]
    \decljudgement{\xvdash{prop}{activ} \var{p} \Downarrow \var{ev}}
    \begin{mathpar}
   \inferrule* [Right=\rulename{PAB}]
       {  }
       { \xvdash{prop}{activ} (\smabinding{x}{\var{lhs}}{a}{\var{rhs}})_{\vroot_x} \Downarrow \eactivate{\vroot_x}}
       \\
   \inferrule* [Right=\rulename{PAC$_1$}]
       { }
       { \xvdash{prop}{activ} (\smacomponent{a}{x}{\var{ps}})_{\vroot_x} \Downarrow \eactivate{\vroot_x.x} }
    \and
   \inferrule* [Right=\rulename{PAC$_2$}]
       { \xvdash{prop}{activ} (\var{ps}_i)_{\vroot_x.x} \Downarrow \var{ev} }
       { \xvdash{prop}{activ} (\smacomponent{a}{x}{\var{ps}})_{\vroot_x}
         \Downarrow \var{ev} }
    \end{mathpar}
    \decljudgement{\xvdash{prop}{deactiv} \var{p} \Downarrow \var{ev}}
    \begin{mathpar}
   \inferrule* [Right=\rulename{PDB}]
       {  }
       { \xvdash{prop}{deactiv} (\smabinding{x}{\var{lhs}}{_}{\var{rhs}})_{\vroot_x} \Downarrow \edeactivate{\vroot_x.x}}
       \and
   \inferrule* [Right=\rulename{PDC$_1$}]
       { }
       { \xvdash{prop}{deactiv} (\smacomponent{_}{x}{\var{ps}})_{\vroot_x} \Downarrow \edeactivate{\vroot_x.x} }
    \and
   \inferrule* [Right=\rulename{PDC$_2$}]
       { \xvdash{prop}{deactiv} (\var{ps}_i)_{\vroot_x.x} \Downarrow \var{ev} }
       { \xvdash{prop}{deactiv} (\smacomponent{_}{x}{\var{ps}})_{\vroot_x}
         \Downarrow \var{ev} }
       \end{mathpar}
    \decljudgement{E, A, T, E' A' \xvdash{prop}{proc} \var{p} \Downarrow \var{ev}}
    \begin{mathpar}
   \inferrule* [Right=\rulename{PC$_1$}]
        { \smafindproc{\var{p}}{\vroot_x.x}{\smacomponent{a}{x}{\var{ps}}} \\
          \eactivate{\vroot_x.x} \in T \\
          \xvdash{prop}{activ} (\var{ps}_i)_{\vroot_x.x} \Downarrow \eactivate{y}
        }
        { E, A, T, E', A' \xvdash{prop}{proc} \var{p} \Downarrow \eactivate{y} }
        \and
    \inferrule* [Right=\rulename{PC$_2$}]
        { \smafindproc{\var{p}}{\vroot_x.x}{\smacomponent{_}{x}{\var{ps}}} \\
          \edeactivate{\vroot_x.x} \in T \\
          \xvdash{prop}{deactiv} (\var{ps}_i)_{\vroot_x.x} \Downarrow \edeactivate{y}
        }
        { E, A, T, E', A' \xvdash{prop}{proc} \var{p} \Downarrow
          \edeactivate{y} }
  \end{mathpar}
    \caption{Propagation rules for components}
    \label{fig:rules-prop-2}
\end{figure}

The last set of rules, in \cref{fig:rules-prop-2}, describes how
components propagate events to their children.
Rule \rulename{PC$_1$} specifies the propagation of activation events, while
rule \rulename{PC$_2$} specifies the propagation of deactivation events.
Each is specified by a judgment
$\xvdash{prop}{(de)activ} p \Downarrow \var{ev}$, which follows the same logic
as the judgment for initial activation.
The only difference between these two sets of rules is that
activation is only propagated to children for components/bindings marked with
\smainit{a}, while deactivation is always propagated.

\subsubsection{Safe reactions}
The propagation rule by itself is not enough to ensure the right semantics is
given to propagation of events. Indeed, consider the pair of processes
\lstinline[language=smalite]{a: 0 =: y; (x/y > 10) -> t} and suppose that
assignment \texttt{a} is triggered.
According to our semantics, it generates an event
\eassign{\texttt{0}}{\texttt{y}}.
In turn, this triggers the conditional binding, as the value of \texttt{y} as
changed.
However, the behavior of \lstinline[language=smalite]{x/y} is undefined because
it divides by 0.
This means that the predicate for expression evaluation does not apply, and therefore
the premises of the rule for conditional left-hand side activation does not
hold.
In the end, our semantics say that this program evaluates fine, but does not
trigger \texttt{t}.
This is wrong: actually, if we compile and/or execute this program, it will
crash.
Therefore, this program should not admit a semantics at all.
To correct for these cases, we introduce an additional premise in \cref{fig:rule-react},
$E, T, E', A' \xvdash{safe}{proc} \var{p}_{\vroot}$, which ensures
that all expressions that need to be evaluated in the process \var{p}
rooted at \vroot{} evaluate without undefined behavior.

The definition of this judgment is presented in \cref{fig:rules-safe}.
Execution of a process that does not propagate events (property, \kw{Spike}) is
always safe.
An assignment is safe if, when triggered, the evaluation of its expression
is safe (rule \rulename{SA}).
Both bindings and components are safe if they are inactive (rules
\rulename{SB$_1$} and \rulename{SC$_1$}).
If an assignment is active, then its left-hand side must be safe:
if the left-hand side is
a condition, either there is no assignment to its free variables, in which case
it does not need to be evaluated (rule \rulename{SLI$_1$}), or it evaluates to a
boolean value (rule \rulename{SLI$_2$}).
An active component is safe if all its children processes are safe (rule \rulename{SC$_2$}).

\begin{figure}[t]
  \decljudgement{E, T, E' \xvdash{safe}{lhs} \var{lhs}}
  \begin{mathpar}
    \inferrule* [Right=\rulename{SLT}]
        { }
        { E, T, E' \xvdash{safe}{lhs} x }
    \and
    \inferrule* [Right=\rulename{SLA}]
        { }
        { E, T, E' \xvdash{safe}{lhs} \lhsactivate{x} }
    \and
    \inferrule* [Right=\rulename{SLD}]
        { }
        { E, T, E' \xvdash{safe}{lhs} \lhsdeactivate{x} }
    \and
    \inferrule* [Right=\rulename{SLC}]
        { }
        { E, T, E' \xvdash{safe}{lhs} \lhschange{x} }
    \and
    \inferrule* [Right=\rulename{SLI$_1$}]
        { \forall x \in \mathsf{free}(e),\ \forall v,\ \eassign{v}{x} \notin T }
        { E, T, E' \xvdash{safe}{lhs} \lhscond{e} }
    \and
    \inferrule* [Right=\rulename{SLI$_2$}]
        { E, E' \xvdash{exp}{} e \Downarrow v \\ v \in \set{\typ{true},\typ{false}} }
        { E, T, E' \xvdash{safe}{lhs} \lhscond{e} }
  \end{mathpar}
  \decljudgement{E, T, E', A' \xvdash{safe}{proc} \var{p}_{\vroot{}}}
  \begin{mathpar}
    \inferrule* [Right=\rulename{SP}]
        {  }
        { E, T, E', A' \xvdash{safe}{proc} (\smaprop{\var{ty}}{x}{e})_{\vroot_x} }
    \and
    \inferrule* [Right=\rulename{SS}]
        {  }
        { E, T, E', A' \xvdash{safe}{proc} (\smaspike{x})_{\vroot_x} }
    \and
    \inferrule* [Right=\rulename{SA}]
        { \etrigger{\vroot_x.x} \in T \implies E, E' \xvdash{exp}{} e \Downarrow v }
        { E, T, E', A' \xvdash{safe}{proc} (\smaassign{x}{e}{y})_{\vroot_x} }
        \\
    \inferrule* [Right=\rulename{SB$_1$}]
        { \vroot_x.x \notin A' }
        { E, T, E', A' \xvdash{safe}{proc} (\smabinding{x}{\var{lhs}}{_}{\var{rhs}})_{\vroot_x} }
    \and
    \inferrule* [Right=\rulename{SB$_2$}]
        { E, T, E' \xvdash{safe}{lhs} \var{lhs} }
        { E, T, E', A' \xvdash{safe}{proc} \smabinding{x}{\var{lhs}}{_}{\var{rhs}} }
    \and
    \inferrule* [Right=\rulename{SC$_1$}]
        { \vroot_x.x \notin A' }
        { E, T, E', A' \xvdash{safe}{proc} (\smacomponent{_}{x}{\var{ps}})_{\vroot_x} }
    \and
    \inferrule* [Right=\rulename{SC$_2$}]
        { \forall i,\ E, T, E', A' \xvdash{safe}{proc} \var{ps_i}_{\vroot_x.x} }
        { E, T, E', A' \xvdash{safe}{proc} (\smacomponent{_}{x}{\var{ps}})_{\vroot_x} }
  \end{mathpar}
  \caption{Safe reaction rules}
  \label{fig:rules-safe}
\end{figure}

\subsubsection{State update}

The two judgments described above specify what events are emitted during the
reaction. It remains to specify how the state of the reactive system is updated
by these events. This is the role of the judgment
$E, A \xvdash{update}{} T \Downarrow E', A'$, which specifies that
``when it receives events T, the state of a process updates from
(E, A) to (E', A')''.
The unique rule specifying this judgment is presented in
\cref{fig:rule-update}.

Its first two premises specify the updated environment $E'$: for each path $p$, either
there is an \kw{Assign} event to $p$, in which case $E'(x)$ is set to its value, or
there is none, in which case $E'(x)$ keeps the same value as $E(x)$.
The first premise implies a strong constraint of \smalite{} programs, that was not
explicit until now: two assignments to the same property with different values
may not occur during the same reaction.
This is a stronger constraint than was chosen in the original implementation of
\smala{} (where the value of the ``final'' assignment was kept), but it
simplifies the semantic model and facilitates the generation of simple and
efficient imperative code.
In practice, this semantic constraint is satisfied for any program that respects
a \ac{RSSA} property: each reaction only contains one assignment to a given
property.
This property is checked by our prototype compiler.

The remaining premises specify the updated activation $A'$: an \eactivate{p}
event adds $p$ in $A'$, a \edeactivate{p} event removes it.
If neither type of events are emitted, then $p$ keeps its activation status.
These premises imply a second constraint: \eactivate{p} and
\edeactivate{p} events may not occur during the same reaction.
Like the previous constraint, this choice simplifies both the semantics
and compilation scheme, and is checked statically during compilation.

\begin{figure}
  \decljudgement{E, A \xvdash{update}{} T \Downarrow E', A'}
  \begin{mathpar}
    \inferrule* [Right=\rulename{U}]
        { \forall p\ v,\ \eassign{v}{p} \in T \implies E'(p) = \lfloor v \rfloor \\
          \forall p,\ (\forall v,\ \eassign{v}{p} \notin T) \implies E'(p) = E(p) \\
          \forall p,\ \eactivate{p} \in T \implies p \in A' \\
          \forall p,\ \edeactivate{p} \in T \implies p \notin A' \\
          \forall p,\ (\eactivate{p} \notin T \land \edeactivate{p} \notin T)
          \implies (p \in A' \iff p \in A)
        }
        { E, A \xvdash{update}{} T \Downarrow E', A' }
  \end{mathpar}
  \caption{State update rule}
  \label{fig:rule-update}
\end{figure}

\subsubsection{Discussion}
Going back to the rule that composes them in \cref{fig:rule-react}, we can now
get a high-level view of how these three judgments interact interact to specify
how a \smalite{} process reacts to an event.
As discussed, the \var{prop} and \var{safe} rules specify which events are
produced, while the \var{update} rule specifies how state is updated.
These definitions appear to be mutually recursive.
On the one hand, in \var{prop}, the updated state $E', A'$ is used as argument to compute generated events.
On the other hand, in \var{update}, the set of events $T$ is used as an argument
to compute the updated state $E', A'$.
To make these semantics executable, one would most likely need to implement them
as a pair of mutually-defined fixed-points.

\section{Conclusion and Future work }
\label{sec:future}

In this paper, we have presented \smalite{}, a minimal language capable of encoding more fully-featured \acp{UIDL} such as \smala{}.
The denotational semantics of the language, along with a prototype compiler, have been
implemented in the Rocq prover:
this is a first stepping stone towards a verification framework for UIDLs in general and \smala{} in particular.
We propose here some leads for future work.

\subsection{A formally verified compiler for \smalite{}}
\label{sec:verified-compilation}
We have implemented a prototype compiler for \smalite{} programs that generates C
code which can be linked against the SDL library~\cite{SDL3} to define reactive \ac{GUI}
programs.
This compiler generates executable code for \smalite{} programs by finding
a static scheduling of instructions that implement events (setting the value of
property, (un)setting activation flag), guarded by conditions that
emulate the left-hand side of bindings.
This involves three major passes between separate \acp{IR}:
\begin{enumerate}
\item Explicitly compute the event propagation graph of the program, where
  vertices are instructions and edges are guarded by conditions that correspond
  to the left-hand side of bindings
\item Checking that this graph does not contain any cycle, and flattening it by
  transforming its transitive closure into direct associations between external events and lists
  of guarded instructions
\item For each external event, schedule the guarded instructions according to
  dependencies (e.g. an assignment to \texttt{x} must be processed before an
  instruction that depends on \texttt{x}), and generate a function that
  implements the reaction to the event
\end{enumerate}

Our prototype compiler targets the Obc \ac{IR} of the \velus{}
compiler~\cite{Brun:PhD:2020}.
Obc is an imperative object-oriented language where each class has fields and
methods.
In our compiler, we use fields to store the values of properties and the
activation state of permanent processes, and generate one method for each external
event.
Then, we reuse the Obc-to-Clight pass of \velus{} to produce a Clight program.
Clight is one of the frontend language of the CompCert verified
compiler~\cite{Leroy:Realistic:2009} on which \velus{}, and therefore our prototype
compiler, are based.

In the future, we hope to reuse the correctness proofs of \velus{} and CompCert to
build an end-to-end semantics preservation proof from the denotational semantics
presented in this paper down to the assembly semantics provided by CompCert.
The missing piece is, of course, a correctness proof that relates our
denotational semantics for \smalite{} to the operational semantics of
Obc~\cite[\textsection{}4.1.2]{Brun:PhD:2020}.
We expect this proof to be difficult, for three reasons.
First, our denotational semantics model is clearly more abstract than the
operational semantics used for Obc: the former asserts the existence of
objects on which a set of relations hold, while the latter describes precisely
how these objects are computed.
Second, the correctness of our compilation scheme heavily depends on the
well-formedness of the source program (absence of dependency loops, absence of
contradictory assignments, etc.) which might be difficult to specify precisely
and reason about.
Last, the rule for reactions described in \cref{fig:rule-react} relies on a
complex predicate to define the content of the set of events $T$.
We are afraid that reasoning about this predicate in semantics preservation
proofs will require proving implications in two directions (source-to-target and
target-to-source), which might incur a significant amount of work for each
compilation pass.

\subsection{Static analysis, simplifications and optimizations}

Our prototype compiler is, in some respects, very naïve, and future work will
focus on improving it so that it generates more efficient
imperative code, and accepts a larger set of source \smalite{} programs.

Indeed, compilation to imperative code with a fixed schedule places a restriction on which programs
may be accepted: some programs with well-defined and deterministic semantics may
be rejected during compilation.
For instance, consider a program with only two bindings: \lstinline[language=smalite]{x -> y; y -> x}.
In theory, this program is not schedulable, as there is a dependency loop
between the triggering of \texttt{x} and \texttt{y}.
In practice however, the semantics of this program are clear: either \texttt{x}
and \texttt{y} are both triggered, or they both are not.
To generate imperative code, we need to somehow cut the
dependency cycle, but it is not obvious how to do so in general.

Another, more complex loop is actually showcased in our example, on
\texttt{lines 29-30}.
\texttt{Line 29} specifies that releasing \texttt{btn2} sets \texttt{count} back to \texttt{3}.
\texttt{Line 30} specifies that changing \texttt{count} triggers a test that activates
\texttt{btn2} if \texttt{count < 3}.
How should these two bindings be scheduled?
On the one hand, before checking the first binding, it must be determined whether or
not \texttt{btn2} is active, so \texttt{line 30} should be scheduled before
\texttt{line 29}.
On the other hand, \texttt{line 29} sets \texttt{count}, so it should be scheduled
before \texttt{line 30} that listens to a change on \texttt{count}.
This looks like a dependency loop, but on closer inspection the second
scheduling is never useful: indeed, \texttt{line 29} sets \texttt{count} to
\texttt{3}, therefore, \texttt{count < 3} will never be true on a cycle where this
assignment is activated.
A general way to filter this type of false dependency loops would be to ``cut''
chains of bindings that include conditions that will never evaluate to true.
To do so, we could use some type of static analysis such as abstract
interpretation~\cite{Cousot:StaticDynamic:1976}.
Moreover, statically simplifying conditions and cutting useless bindings
would also make the generated code more efficient.

\subsection{Extending \smalite{}}
The example presented in \cref{sec:example} could be simplified with a few higher-level constructs.
First, we see a very similar code pattern of declaring an assignment and having
a unique binding to that assignment at several places
(\texttt{lines 14, 15, 17, 29, ...}).
These could be simplified with a meta-process that encodes ``binding to an
assignment''.
In \smala{}, this is implemented by assignment sequences, with syntax
\lstinline|press -> { 255 =: green }|.
This feature could be added as part of a more general source language than the
one we present in this paper, and compiled down to the simpler constructs of \smalite{}.

Another unwelcome repetition appears in the definitions of the buttons
\texttt{btn1} and \texttt{btn2}, which are essentially identical bar a few
parameter.
The full \smala{} language allows the user to define parameterized components which
can then be instantiated in more complex programs.
It is not yet clear what would be the best way to treat such a feature in our
formalization: either user-defined components could be inlined into a single
\smalite{} program, or they could be compiled separately, making \smalite{} and the
other intermediate representations more complex.

\subsection{Graphical semantics and properties}
The semantic model we propose in this paper describes how the internal state of
a program is updated in reaction to an event.
It does not specify how this internal state is related to the observable
behavior of the program.
In particular, all the components used in the example (\kw{Frame},
\kw{Rectangle}, etc) have graphical semantics: their activation, and the
value of their property affects what is displayed on the screen.
Furthermore, the events that are modeled by spikes (\texttt{close},
\texttt{released}, etc) correspond to user actions.
The structure of graphical components could be formalized as a scene graph, while
the behavior of interactions could be specified using low-level events (click on
the mouse at specific coordinates, etc).
Modeling these graphical and interactive semantics at the level of the \smala{}
source language would facilitate two high-level goals.

First, mechanize a compilation correctness proof for the code that binds the
reactive program to the system libraries that implement user interactions; for
our prototype compiler, that would be SDL3.
Writing this proof would first require axiomatizing the behavior of all the SDL3
functions in use: drawing functions would affect the scene graph, while
event-listening functions would be related to low-level events.

Second, it would be possible to reason on graphical properties of the system.
Indeed, many safety-critical interactive systems, such as airplane cockpit
\acp{GUI} are bound by strict norms.
For instance, the ED 143~\cite{ED143} specifies the behavior of the \ac{TCAS}, which
prevents aircraft from crashing into each other by alerting the pilot of
imminent collisions and requesting altitude change.
In particular, it specifies how incoming aircraft should be displayed on the screen.

In~\cite{Prun:GraphicalProperties:2022}, the authors present a technique to
formally verify that this specification holds for a given implementation, by
generating \acp{WP} and verifying them using the Z3 SMT
Solver~\cite{demouraZ3EfficientSMT2008}.
We could ground these results by proving the correspondence of the \ac{WP}
generation algorithm with our semantics model, and using a tool such as
SMTCoq~\cite{Armand:SMTCoq:2011} to transport the proof of correctness generated
by Z3 into Rocq logic.
This would give us a mechanized Rocq proof that these properties hold for the
source program and, by applying the compiler correctness proof mentioned in
\cref{sec:verified-compilation}, that they hold for the generated binary.

Another approach would be to reuse the work proposed in
\cite{Nalpon:UIDL:2022,Nalpon:PhD:2023}, which proposes a denotational semantic model based on
bigraphs for a \ac{UIDL} very similar to ours.
It then uses the PRISM model checker~\cite{PRISM:2011} to prove that the
resulting bigraphical reactive system implies graphical properties of interest.
To take advantage of these results, we would first need to prove that our own
relational semantic model is equivalent to the bigraphical model.

\printbibliography{}

\end{document}